%% file: firobf.tex
\documentclass[conference]{IEEEtran}

\usepackage{url}
\usepackage{array}
\usepackage{xcolor}
\usepackage{cancel}
\usepackage{siunitx}
\usepackage{multirow}
\usepackage{graphicx}
\usepackage{textcomp}
\usepackage{multirow}
\usepackage{stfloats}
\usepackage{algorithm}
\usepackage[T1]{fontenc}
\usepackage[noend]{algpseudocode}
\usepackage[10pt]{moresize}
\usepackage[utf8]{inputenc}
\newif{\ifusebiblatex}
\usebiblatexfalse
\ifusebiblatex
\usepackage[
style=ieee, %use ieee style as we have a ieee paper
backend=biber, %use modern biber instead of old natbib
isbn=false, %just drop ISBNs
doi=false, %just drop DOIs
eprint=false, %just drop anything in the eprint value
maxcitenames=2, %when using \textcite, at max print 2 author names
mincitenames=1, %when using \textcite and we have many authors (et al), print 1 author name then et al
minbibnames=1, %in bibliography, when there are many authors (et al), print 1 author name then et al
maxbibnames=6 %in bibliography, print at maximum 6 author names
]{biblatex}
\DeclareSourcemap{
	\maps[datatype=bibtex]{
		\map{
			\step[fieldset=abstract, null]
		}
		\map[overwrite]{
			\step[fieldsource=shortjournal]
			\step[fieldset=journaltitle,origfieldval]
		}
	}
}
\AtEveryBibitem{%
	\ifentrytype{article}{%
		\clearfield{url}%
		\clearfield{urldate}%
		\clearfield{location}%
		\clearfield{address}%
		\clearfield{publisher}%
		\clearfield{editor}
	}{}
	\ifentrytype{inproceedings}{%
		\clearfield{url}%
		\clearfield{urldate}%
		\clearfield{location}%
		\clearfield{publisher}%
		\clearfield{address}%
		\clearfield{editor}
	}{}
	\ifentrytype{book}{%
		\clearfield{url}%
		\clearfield{urldate}%
		\clearfield{location}%
		\clearfield{publisher}%
		\clearfield{address}%
	}{}
}
\DeclareFieldFormat{url}{\url{#1}}
\DeclareFieldFormat{titlecase}{#1}
\DeclareFieldFormat{sentencecase}{#1}

\defbibheading{References}{\section*{\refname}\addcontentsline{toc}{section}{\refname}}

\makeatletter

\AtBeginBibliography{\vskip 0.3\baselineskip plus 0.1\baselineskip minus 0.1\baselineskip}
\makeatother
\usepackage{xpatch}

\xpatchbibdriver{online}
{\printtext[parens]{\usebibmacro{date}}}
{\iffieldundef{year}{}{\printtext[parens]{\usebibmacro{date}}}}
{}{}

\usepackage[autostyle=true]{csquotes}

\else
\usepackage[noadjust]{cite}
\fi

\usepackage{mathtools, cuted}
\usepackage{amsmath,amssymb,amsfonts}

\IEEEoverridecommandlockouts

\graphicspath{{./figs/}}

\newcommand{\comment}[1]{}
\newlength{\textfloatsepsave} 
\begin{document}
	
	\title{Hardware Obfuscation of Digital FIR Filters}
	\author{
		\IEEEauthorblockN{Levent~Aksoy\IEEEauthorrefmark{2},
			%Michaela Brunner\IEEEauthorrefmark{3}, 
			Alexander Hepp\IEEEauthorrefmark{3},
			Johanna Baehr\IEEEauthorrefmark{3} and
			Samuel Pagliarini\IEEEauthorrefmark{2}}
		\IEEEauthorblockA{\IEEEauthorrefmark{2}Department of Computer Systems, Tallinn University of Technology, Tallinn, Estonia}
		\IEEEauthorblockA{\IEEEauthorrefmark{3}Department of Electrical and Computer Engineering, Technical University of Munich, Munich, Germany}
	}
	\maketitle
	\begin{abstract}
		A finite impulse response (FIR) filter is a ubiquitous block in digital signal processing applications. Its characteristics are determined by its coefficients, which are the intellectual property (IP) for its designer. However, in a hardware efficient realization, its coefficients become vulnerable to reverse engineering. This paper presents a filter design technique that can protect this IP, taking into account hardware complexity and ensuring that the filter behaves as specified only when a secret key is provided. To do so, coefficients are hidden among decoys, which are selected beyond possible values of coefficients using three alternative methods. As an attack scenario, an adversary at an untrusted \mbox{foundry is} considered. A reverse engineering technique is developed to find the chosen decoy selection method and explore the potential leakage of coefficients through decoys. An \mbox{oracle-less} attack is also used to find the secret key. Experimental results show that the proposed technique can lead to filter designs with competitive hardware complexity and higher resiliency to attacks with respect to previously proposed methods.
	\end{abstract}
	
	\begin{IEEEkeywords}
		hardware obfuscation, IP protection, reverse engineering, oracle-less attack, digital FIR filter design.
	\end{IEEEkeywords}
	
	\input{introduction}
	\input{background}
	\input{architectures}
	\input{reveng}
	\input{results}
	\input{conclusions}
	\input{acknowledgment}
	
	\ifusebiblatex
	\printbibliography[heading=References]
	\else
	\bibliographystyle{IEEEtran}
	\bibliography{firobf_shortened}
	\fi
	
\end{document}

%Email: \{levent.aksoy, samuel.pagliarini\}@taltech.ee
%Email: \{johanna.baehr, michaela.brunner, alex.hepp, sigl\}@tum.de}

%% file: introduction.tex
\section{Introduction}

\begin{figure}[t]
	\centerline{\includegraphics[width=6.0cm]{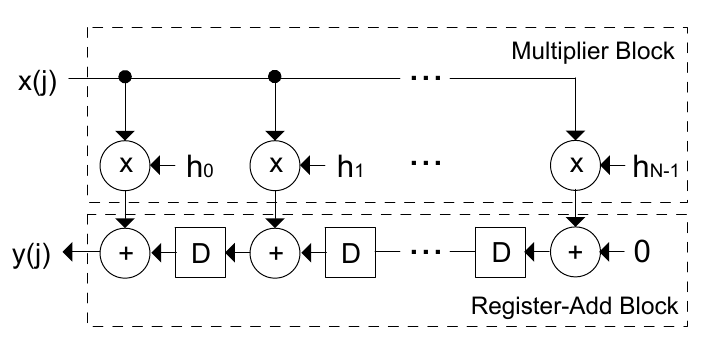}}
	\vspace*{-3mm}
	\caption{Parallel transposed form FIR filter design.}
	\vspace*{-7mm}
	\label{fig:fir-trans}
\end{figure}

Digital filtering is frequently used in digital signal processing (DSP) applications and finite impulse response (FIR) filters are generally preferred due to their stability and linear phase property~\cite{wanhammar99}. The output of an FIR filter $y(j)$ is equal to $\sum_{i=0}^{N-1}h_i\cdot x(j-i)$, where $N$ is the filter length, $h_i$ is the $i^{th}$ filter coefficient, and $x(j-i)$ is the $i^{th}$ previous filter input with $0 \leq i \leq N-1$. Fig.~\ref{fig:fir-trans} shows the design of transposed form FIR filter. Note that since coefficients determine the filter behavior, they are actually an intellectual property (IP).

In filter design, on one hand, coefficients can be \mbox{stored in} memory, revealing no information to an adversary at an untrusted foundry. Due to the usage of memory and multipliers, which take two inputs as variables, the hardware complexity is increased in this case. On the other hand, since coefficients are determined beforehand, multipliers, whose one input is a variable and the other is a constant, can be used, enabling high-level and logic synthesis tools to embed coefficients into hardware and reduce its complexity further~\cite{aksoy14_tutorial}. In this case, coefficients become vulnerable. For example, after resetting registers and applying a constant 1 value to the filter input, the filter output $y(j)$ is computed as $\sum_{i=0}^{j} h_i$ in the first $N-1$ clock cycles. Thus, by observing the filter output in $N-1$ clock cycles, each coefficient can be determined as follows:
\begin{equation}
	h_i =
	\begin{cases}
		y(0),        & \text{if $i=0$}\\
		y(i)-y(i-1), & \text{otherwise.}
	\end{cases}
	\label{eqn:coefs}
\end{equation}

Over the years, many efficient techniques have been introduced for the protection of IPs. Among these techniques, logic locking methods~\cite{dupuis19} offer a protection against a diverse array of adversaries~\cite{yasin17}. They insert additional logic at \mbox{gate-level}, which is driven by a key, so that the circuit behaves as expected only when the secret key is applied. Moreover, \mbox{high-level} obfuscation techniques have been proposed to protect IPs~\cite{islam20, pilato21,  aksoy21}. In~\cite{islam20}, the whole arithmetic function is obfuscated rather than only constants. In~\cite{pilato21}, constants can be obfuscated by simply replacing their bits by key inputs which are stored in a memory. The obfuscation technique of~\cite{aksoy21} hides filter coefficients behind decoys. However, since there may exist many possible sets of coefficients that satisfy the filter specification~\cite{aksoy14}, all these methods cannot ensure that the encrypted filter exhibits the specified behavior only when the secret key is provided. To the best of our knowledge, only the approach of~\cite{lao15} can ensure this property using \mbox{high-level} transformations, a \mbox{key-based} finite-state machine, and a reconfigurator. 

In this paper, we introduce a filter design technique which obfuscates a digital FIR filter based on the given filter specification. In this technique, the filter behavior is formulated as a linear programming (LP) problem and the filter coefficients, that respect the given filter specification, and their lower and upper bounds are computed. For each filter coefficient, decoy(s) are chosen beyond the lower and upper bounds of the associated coefficient using three alternative decoy selection methods (DSMs). As done in~\cite{aksoy21}, coefficients are protected by hiding them among these decoys using an additional logic with key inputs. This technique can be applied to any type of FIR filter and it works at \mbox{register-transfer} level, enabling logic synthesis tools to reduce the hardware complexity. To find the resiliency of the obfuscated design against an attack at an untrusted foundry, a reverse engineering method based on machine learning is developed to find the DSM and discover possible coefficients among decoys. Also, an \mbox{oracle-less} attack~\cite{alaql21} is used to find the secret key. It is shown that the proposed technique using a specific DSM leads to filters with a hardware complexity similar to those obtained by previously proposed methods. However, the attacks are not successful on these proposed designs, while they can discover some coefficients of filters encrypted by other methods. 

The remainder of this paper is organized as follows: Section~\ref{sec:background} gives background concepts. The proposed filter design technique is introduced in Section~\ref{sec:obf}. Section~\ref{sec:reveng} describes the developed reverse engineering technique. Experimental results are given in Section~\ref{sec:results}. Finally, Section~\ref{sec:conclusions} concludes the paper.

%The proposed technique realizes the filter under the folded architecture~\cite{parhi99}.

%and obfuscates the fundamental \mbox{time-multiplexed} constant multiplication (TMCM) block which realizes the multiplication of the filter input by a constant selected among all the filter coefficients at a time.

%For example, filters locked by logic locking techniques~\cite{yasin16, yasin17, xie19}, that are resilient to powerful logic locking attacks, such as the satisfiability (SAT)-based attack~\cite{subramanyan15}, act differently only on a small number of outputs under a wrong key, leading to a filter behavior very close to the original one.

%Moreover, these attacks are applied to filters locked by conventional techniques~\cite{roy08, yasin16} and obfuscated by the technique of~\cite{aksoy21}.

%The impact of decoy selection methods on the hardware complexity and resiliency to the attacks is explored.

%Furthermore, the behavior of filters obfuscated by the proposed technique violates the filter specification whenever a wrong key is provided. 

%However, in this case, In the filter design, since its coefficients are determined beforehand, while one input of the multiplier is a variable, the other is a constant. This enables high-level and logic synthesis tools to reduce the hardware complexity further~\cite{aksoy14_tutorial}, which is not possible if coefficients are stored in a memory and multipliers, that take two variables as inputs, are used. However, coefficients become vulnerable to an adversary at an untrusted foundry in this case. 

%% file: background.tex
\section{Background}
\label{sec:background}

\subsection{Filter Design}
\label{subsec:fdo}

The filter design problem can be defined as finding filter coefficients that satisfy filter constraints based on the  filter specification \textit{fspec}, given as a five-tuple, i.e., filter length $N$, passband $w_p$ and stopband $w_s$ frequencies, and passband $\delta_p$ and stopband $\delta_s$ ripples~\cite{aksoy14}. The zero-phase frequency response (ZPFR) of a symmetric FIR filter is given as\footnote{The frequency response of an asymmetric filter can be found in~\cite{lim90}.},
\begin{equation}
	G(w)=\sum_{i=0}^{\lfloor M \rfloor}e_ih_icos(w(M-i)),
	\label{eqn:zpfr}
\end{equation}
where $M = (N-1)/2$, $h_i \in \mathbb{R}$ with $-1 \leq h_i \leq 1$, $w \in \mathbb{R}$ is the angular frequency, and $e_i = 2 - K_{i,M}$ with $K_{i,M}$ is the Kronecker delta\footnote{The $K_{a,b}$ function is 1 when $a$ is equal to $b$. Otherwise, it is 0.}. Considering a low-pass filter and assuming that the pass-band and stop-band gains are set to 1 and 0, respectively, the filter must satisfy the following constraints:
\begin{equation}
	\begin{array}{rll}
		1-\delta_p &\hspace*{-2mm}\leq G(w) \leq 1+\delta_p, & w \in [0,w_p] \\
		-\delta_s  &\hspace*{-2mm}\leq G(w) \leq \delta_s, & w \in [w_s,\pi].
	\end{array}
	\label{eqn:firchar-ge}
\end{equation}

The filter design problem can be formulated as an LP problem, for which there exists a polynomial-time algorithm~\cite{dantzig-book}. 

%Observe that there can be many possible sets of coefficients, each satisfying the filter constraints.

%Since the pass-band gain is not relevant for many DSP applications, it can be compensated in the filter design by adding a scaling factor into the filter constraints as a continuous variable~\cite{saramaki01, aksoy14}.

%In DSP applications, it may be desirable to minimize the peak weighted ripple~\cite{samueli89}, the normalized peak ripple (NPR)~\cite{yu10}, or the NPR magnitude~\cite{xu07}. 

%A straightforward filter design technique (SFDT) follows two steps: i)~given \textit{fspec}, find the filter coefficients, that respect the filter constraints, using a filter design method, such as windowing~\cite{window-method}, McClellan-Parks-Rabiner algorithm~\cite{mpr-method}, or linear programming~\cite{lp-method}; ii)~implement the filter under the given architecture based on the computed filter coefficients.

%Since there exist many possible sets of coefficients satisfying the filter constraints, filter design optimization algorithms incorporate sophisticated techniques such as local search~\cite{samueli89,xu07} and exhaustive search methods, including branch-and-bound~\cite{saramaki01} and depth-first search~\cite{yu10,shi11, aksoy14} techniques in order to reduce the hardware complexity of the filter. 

\subsection{Folded Implementation of Digital FIR Filters}

After the floating-point filter coefficients are determined, respecting the filter specification, they are converted to integer constants, since floating-point addition and multiplication operations have more hardware complexity than their integer counterparts~\cite{horowitz14}. Also, the folded design architecture, where computing resources are re-used, is preferred to the parallel design architecture, leading to a significant reduction in area, but increasing the latency of the computation~\cite{parhi99}. 

\begin{figure}[t]
	\centerline{\includegraphics[width=7.5cm]{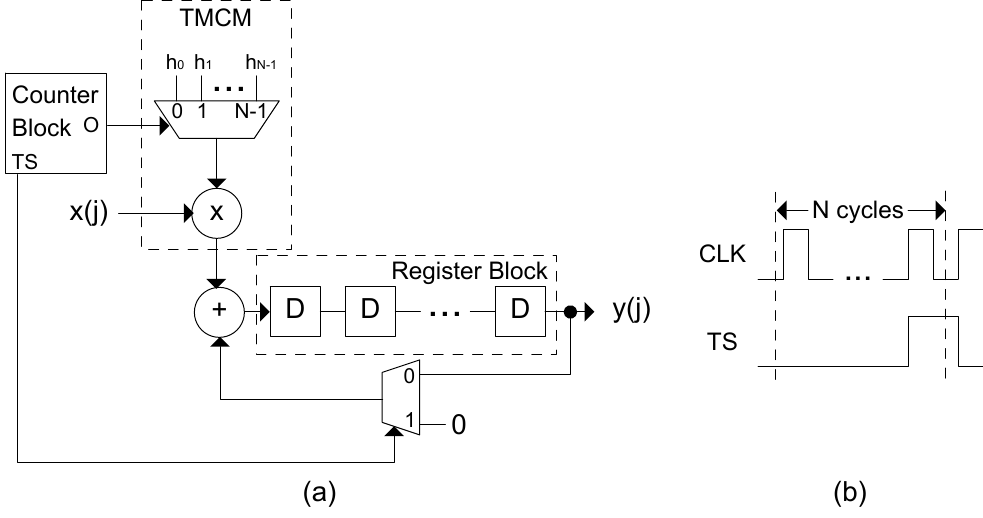}}
	\vspace*{-4mm}
	\caption{(a)~Folded design of the transposed filter; (b)~the timing signal $TS$.}
	\vspace*{-6mm}
	\label{fig:fir-folded}
\end{figure}

Fig.~\ref{fig:fir-folded}(a) shows the folded design of the filter given in Fig.~\ref{fig:fir-trans}, where the \mbox{time-multiplexed} constant multiplication (TMCM) block realizes the multiplication of the filter input by a constant selected among all filter coefficients at a time~\cite{aksoy14_tmcm}. The register block includes $N-1$ cascaded registers whose counterparts in the parallel design are the ones in the register-add block. The $\lceil log_2{N} \rceil$-bit counter counts from 0 to $N-1$ and generates the timing signal $TS$ shown in Fig.~\ref{fig:fir-folded}(b). Note that $CLK$ denotes the clock signal fed to all registers which was not shown in Fig.~\ref{fig:fir-folded}(a) for the sake of clarity. 

In the proposed filter design technique, the TMCM block including filter coefficients is obfuscated using decoys.

%Although the filter design complexity is reduced under the folded architecture by reusing the common operations, the filter output is obtained in $N$ clock cycles. 

%In a folded design, the TMCM operation is the main block which can be realized under different architectures, as shown in~\cite{aksoy14_tmcm}.

\subsection{Attacker Model for Reverse Engineering}

In our threat model, an adversary at an untrusted foundry aims to retrieve coefficients of the FIR filter obfuscated as described in Section~\ref{sec:obf}. In this case, it is reasonable to assume that the adversary can identify the block implementing the TMCM operation, either by inspecting the chip data sheet or by using a fuzzy reverse engineering method which labels netlist partitions with their functionality~\cite{baeh2020}. Thus, the Boolean function of the TMCM block in the netlist is available to the adversary, as well as the implementation details given in Section~\ref{sec:obf}. The adversary is assumed to have no further a priori knowledge about the system, such as the filter specification or an unlocked design, and therefore cannot produce an oracle for the filter design. As a consequence, filter coefficients and decoys represent secret information that must be protected. Hence, there must be no information in the netlist that can be used to discern original coefficients from decoys.

% \subsection{Graph Representation of Netlists}

% To reverse engineer a given netlist, it is common to parse it into a graph representation. The graph $G = (V,E)$ is generated, such that each node $v$ is a standard cell. $V$ is thus a set of standard cells in the respective netlist. An edge $e \in E$ is generated between two standard cells $v_1, v_2 \in V$ if there is a net in the design netlist driven from standard cell $v_1$ and driving $v_2$. As net names and cell names are arbitrary and should not influence the classification, no attributes of $v$ are kept. Placement information is not utilized, thus the edges $e$ have unit length.

%% file: architectures.tex
\section{The Proposed FIR Filter Design Technique}
\label{sec:obf}

In this section, we present an FIR filter design technique that generates an obfuscated filter whose behavior is the same as the specified one only when the secret key is provided. Its steps are described in detail in the following subsections.

\subsection{Finding Filter Coefficients and Decoys}

First, based on the \textit{fspec}, the original filter coefficients are computed by solving the constraints given in Eq.~\ref{eqn:firchar-ge}. 

Second, the lower bound of each coefficient $h_i$, where \mbox{$0 \leq i \leq N-1$}, is found by solving the following LP problem:
\begin{equation}
	\begin{array}{ll}
		\hspace*{-3mm} minimize: cf = h_i & \\
		\hspace*{-3mm} subject~to: 1-\delta_p \leq G(w) \leq 1+\delta_p, & w \in [0,w_p] \\
		\hspace*{-3.5mm} ~~~~~~~~~~~~~~~~~-\delta_s \leq G(w) \leq \delta_s, & w \in [w_s,\pi] \\
		\hspace*{-3mm} ~~~~~~~~~~~~~~~~~~~~~~h^l \leq h \leq h^u,
	\end{array}
	\nonumber
\end{equation}
where $cf$ is the cost function and $h^l$ and $h^u$ are the lower and upper bounds of all filter coefficients which were initially assigned to -1 and 1, respectively. The value of $h_i$ in the LP solution corresponds to its lower bound $h_i^l$ and is stored in a set $H^l$. Similarly, the upper bound of each coefficient $h_i^u$ is found when $cf {=} {-}h_i$ and is stored in a set $H^u$. 

%Recall that an LP problem can be solved in polynomial time~\cite{dantzig-book}. For symmetric filters, the number of LP problems to be solved is $2\lfloor M \rfloor+4$. Recall that an LP problem can be solved in polynomial time~\cite{dantzig-book}.

Third, the floating-point coefficients and their lower and upper bounds are converted to integers by multiplying them by $2^Q$, where $Q$ is the quantization value, and finding the least integer greater than or equal to this multiplication result. 

%Note that $Q$ can be a user-defined constant or can be determined in an iterative manner by increasing its value starting from 0 till all the filter constraints are satisfied~\cite{aksoy14}.

Fourth, given the number of key inputs $p$, for each filter coefficient $h_i$, its decoys are selected beyond its lower and upper bounds as given in Algorithm~\ref{algo:decoyassign}. In its \textit{AssignDecoy} function, decoys are preferred to have the same sign with the related filter coefficient to prevent any leakage of information on the original coefficients and are determined to be unique to generate distinct outputs when a wrong key is applied, increasing the output corruption. We consider three DSMs: i)~\mbox{{\sc dsm-hd}} favors decoys with a Hamming distance very close to the related coefficient; ii)~{\sc dsm-rd} chooses decoys randomly whose bit-widths are close to that of the related coefficient; iii)~{\sc dsm-hdrd}, which is a mixture of \mbox{\sc dsm-hd} and \mbox{\sc dsm-rd}, selects a decoy using {\sc dsm-hd} if the related coefficient is protected by a single decoy, otherwise, chooses decoys using {\sc dsm-rd}. Note that {\sc dsm-hd} is introduced to reduce hardware complexity by increasing the sharing of common operations during logic optimization. However, it may lead to a leakage of coefficients when a coefficient has more than one decoy, as shown in Section~\ref{subsec:vulndsm}. Hence, {\sc dsm-rd} is proposed to prevent such a leakage. {\sc dsm-hdrd} is introduced to take advantages of these methods in hardware complexity and resiliency.

\subsection{Realization of the Obfuscated TMCM Block}

\begin{algorithm}[tb]
	\small
	\caption{Assignment of decoys to filter coefficients}
	\begin{algorithmic}[1]
		\Statex \textbf{Given:} $N$, $H^l$, $H^u$, and $p$ 
		\State $noi = 0$ \Comment{Number of iterations}
		\State $nok = 0$ \Comment{Number of used keys}
		\State $D = \{\}$ \Comment{Set of arrays including decoys}
		\State $nd = \{\}$ \Comment{Set including number of decoys}
		\While{$nok < p$}
		\State $nod = 2^{noi}$ \Comment{Number of decoys to be assigned}
		\For{$i=0$ \textbf{to}  $N-1$}
		\State ($nd_i, D_i$) = AssignDecoy($nod$, $h_i^l$, $h_i^u$, $nd_i, D_i$)
		\State $nok = nok + 1$
		\If{$nok == p$}
		\State \textbf{break}
		\EndIf
		\EndFor
		\State $noi = noi + 1$
		\EndWhile
		\State \textbf{return} ($nd$, $D$)
	\end{algorithmic}
	\label{algo:decoyassign}
\end{algorithm}

\begin{figure}[t]
	\centering
	\vspace*{-4mm}
	\includegraphics[width=4cm]{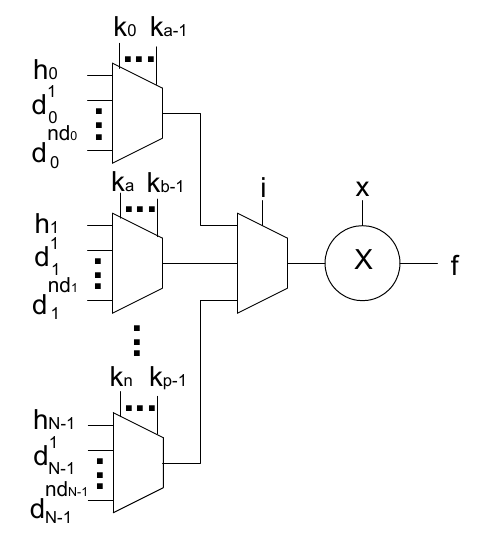}
	\vspace*{-4mm}
	\caption{Obfuscation of the TMCM block.}
	\vspace*{-6mm}
	\label{fig:tmcm-obf}
\end{figure}

Fifth, each filter coefficient $h_i$ is hidden among its decoy(s) $D_i = [d_i^1 \ldots d_i^{nd_i}]$, where $0 \leq i \leq N-1$. To do so, a multiplexor (MUX) is used to select the filter coefficient and its decoy(s) using key input(s) $k$. Locations of a coefficient and its decoys at the MUX inputs are determined randomly. The number of inputs of such a MUX is equal to $nd_i+1$ and the number of its select inputs is $\lceil log_2{(nd_i+1)} \rceil$. Then, another MUX is used to realize the constant multiplication in the TMCM block in the given order using the primary select input $i$ whose bit-width is $\lceil log_2N \rceil$. The number of inputs of this MUX is equal to $N$. Finally, a multiplier is employed to realize the multiplication of the selected constant by the filter input $x$. Its size is equal to $mbw + ibw$, where $mbw$ stands for the maximum bit-width of the filter coefficients and $ibw$ denotes the bit-width of the filter input $x$. Fig.~\ref{fig:tmcm-obf} shows the obfuscated TMCM block. Observe that at least one decoy is multiplied by the filter input under a wrong key. Hence, in a folded design of a filter whose TMCM block is obfuscated using the proposed method, a wrong key always generates a wrong output. Since decoys are selected beyond the possible values of filter coefficients, these wrong outputs lead to a filter behavior other than the specified one.

Finally, the folded FIR filter including the obfuscated TMCM block is implemented as given in Fig.~\ref{fig:fir-folded}(a). A computer-aided design tool is developed to automate the design and verification processes of the obfuscated FIR filter. 

%It takes the filter specification \textit{fspec}, the quantization value $Q$, and the number of key inputs $p$ as inputs. Then, it generates the behavioral description of the obfuscated filter at RTL and the testbench for verification in Verilog. 

%% file: reveng.tex
\section{The Developed Reverse Engineering Method}
\label{sec:reveng}

% missing: how to identify tmcm from FIR filter.
% Describe method, not attacker model
% restructure: put attacker model into background, here rather hav a 1, .2. 3. step by step description. 
% timing, how many training designs, XGBClasifier hyperparams, 

In this section, we describe an attack on the filter design obfuscated as presented in Section~\ref{sec:obf} using reverse engineering. Its steps are described in detail in the following subsections.

\subsection{Identifying the TMCM Block in Obfuscated Filter Design}
\label{subsec:idetmcm}

In the gate-level netlist of an obfuscated FIR filter, its TMCM block can be identified as partitions using a fuzzy reverse engineering method~\cite{baeh2020}. At the end of this step, the attacker has extracted the netlist, i.e., the Boolean function of the TMCM block, that is used in the next step. 

\subsection{Extracting Constants from the TMCM Netlist}
\label{subsec:extcons}

Although constants have been embedded into the TMCM design by the synthesis tool during logic optimization, they can be extracted with reverse engineering. Hypotheses for the constants can be built by using an arbitrary multiplication method as an oracle. The aim is to find constants in a way that the multiplication oracle behaves identically to the reverse engineered TMCM block under all possible filter input $x$ for the given primary select input $i$ and key input $k$. An exhaustive search for constants is not feasible, as it would be necessary to evaluate at most $2^{mbw}$ combinations for all $i$ and $k$ values. We can instead extract constants efficiently with satisfiability (SAT) solving. Let $f_r(i,k,x)$ denote the \mbox{$n$-bit} Boolean function of the TMCM block in the netlist.
% for a given primary select input $i$, key input $k$, and filter input $x$.
We define $f(c,x) = c \cdot x$ to be the \mbox{$n$-bit} Boolean function realizing the multiplication of an $n$-bit constant $c$ by an $n$-bit input $x$. Let $r_{i,k}$ denote any reverse-engineered constant for a particular $i$ and $k$, let $a[0]$ denote the least significant bit (LSBs) of the $n$-bit constant or function $a$ and let $a[0..j]$ denote the range of bits of $a$ with indices from $0$ to $j$. Thus, to determine the list of constants, we search for $r_{i,k}$ that solves the equation
\begin{align*}
	\forall x \quad &f (r_{i,k},x) = f_r(i,k,x).\\
	\intertext{As the LSB of multiplication only depends on the LSBs of $r_{i,k}$ and $x$, we can begin to search for $r_{i,k}[0]$ that satisfies}
	\forall x[0] \quad &f(r_{i,k}[0],x[0])[0] = f_r(i,k,x[0])[0].\\
	\intertext{We subsequently search for $r_{i,k}[j], 0 < j < n$, that satisfies}
	\forall x[0..j]\quad & f(r_{i,k}[0..j],x[0..j])[j]
	= f_r(i,k,x[0..j])[j],
\end{align*}
using the previously calculated $r_{i,k}[0..j-1]$. In this process, the miter logic is constructed as $f \oplus f_r$ and a SAT solver is used to search for an unsatisfying value of $r_{i,k}[j]$ for the miter logic starting from the LSB. Bits of $r_{i,k}$ found in each iteration reduce the search in the next iteration to just one unknown bit. This process is repeated for all values of $i$ and $k$.

Using the SAT formulation described in this subsection, for each filter coefficient $h_i$, we obtain the list of all reverse engineered constants $R_i = [r_{i,1}, \dots, r_{i,nd_i+1}]$ including the coefficient and its decoys. As coefficients and decoys are handled equally by MUXes shown in Fig.~\ref{fig:tmcm-obf}, there is no structural or functional information that allows a discrimination. 

\subsection{Vulnerability of the DSM}
\label{subsec:vulndsm}

Depending on the DSM, an element of $R_i$ may be more probable to be $h_i$. For example, suppose that the original coefficient is 7 $(111)_2$ and 3 decoys, i.e., 6 $(110)_2$, 5 $(101)_2$, and 3 $(011)_2$, are found using {\sc dsm-hd}. Observe that each decoy has a Hamming distance value of 1 to the original coefficient. However, these decoys actually make it obvious that the original coefficient is 7. Because only for this constant, the Hamming distance to all other constants is always 1.

In order to exploit this vulnerability, it is necessary to differentiate the DSM used in an obfuscated TMCM block. We implemented the structural reverse engineering based method described in \cite{baeh2020} and created a library of 7000 synthesized TMCM blocks. In this library, there are 2000 TMCM designs for each DSM and 1000 TMCM designs without decoys. After synthesis, the netlists were parsed from Verilog to a graph representation using pyverilog~\cite{TakamaedaYamazaki2015} and networkx \cite{HSS2008}. Without compromising accuracy, buffers were removed from the netlists as done in \cite{baeh2020}. The graph representation is then converted into a graph embedding using structural properties of the graph as presented in~\cite{baeh2020}. Each graph embedding is labelled with the DSM used to create the netlist or $0$ otherwise. Finally, the XGBClassifier of~\cite{CG2016} is trained without scaling and preprocessing. The XGBClassifier is a time and space efficient implementation of a machine learning tool using gradient tree boosting, that is appropriate for the large number of features and learning samples used in this work. To prove the ability of the machine learning tool to identify the DSM for all designs in the library, we performed a 5-fold cross validation with a test size of 20\% of all TMCM designs. In the cross validation, on average, excellent macro and micro $f1$ performance scores of 0.99 were achieved for the test sets.

\subsection{Finding Original Coefficients}
\label{subsec:foc}

After the DSM used in the TMCM block is classified, the final step is to find $h_i$ in $R_i$. To do so, a script, called \mbox{\sc doc-hd}, is developed to determine $h_i$ based on the observation  explained in Section \ref{subsec:vulndsm}. We note that neither \mbox{\textsc{dsm-rd}} nor \textsc{dsm-hdrd} are vulnerable to this attack. \mbox{\textsc{dsm-rd}} is not vulnerable, as it does not use the Hamming distance based DSM. \textsc{dsm-hdrd} is not vulnerable, as it does use the Hamming distance based DSM only if there is a single decoy, but in this case, it is equally probable that either $r_{i,k}$ in $R_i$ is the filter coefficient $h_i$. This particular observation can be formalized as follows.
%% Example on determination of coefficients through decoys
% For example, suppose that the original coefficient is 7 $(111)_2$ and 3 decoys, i.e., 6 $(110)_2$, and 5 $(101)_2$, and 3 $(011)_2$, were created using {\sc dsm-hd}. Each decoy has a Hamming distance value of 1 to the original coefficient. Observe that these decoys actually make it obvious that the original coefficient is 7, since the Hamming distance of a constant between all other constants is always 1 only for the original coefficient. Based on this observation, a script, called {\sc doc-hd}, is developed to determine the original coefficients through decoys based on the Hamming distance.
To protect against the entire reverse engineering attack, it is important that the attacker cannot discriminate the reverse engineered constants in $R_i$. Thus, decoys must be chosen such that there is no information to tell coefficients and decoys apart. This observation leads to the condition
\begin{align}
	\forall d_i^m~~ P_{\text{\scshape dsm}}[r_{i,k} = h_i] = P_{\text{\scshape dsm}}[r_{i,k} = d_i^m],1 \leq m \leq nd_i\label{eq:securedsm}
\end{align} 
to be held. The probability that the DSM outputs any decoy value must be equal to the probability that the DSM outputs the filter coefficient. In this case, Definition 1 of~\cite{mass17} is fulfilled and the obfuscation scheme becomes provably secure against a reverse engineering based attack \emph{without a priori} knowledge of the filter specification. Eq.~\ref{eq:securedsm} holds for \textsc{dsm-rd}, as decoys are selected from a uniform random distribution. It is equally probable that $h_i$ is sampled from the random distribution by \textsc{dsm-rd}. Eq.~\ref{eq:securedsm} also holds for \textsc{dsm-hdrd}. If there are multiple decoys for a given coefficient, \textsc{dsm-hdrd} is equal to \textsc{dsm-rd}. If there is a single decoy, it is equally probable for both entries in $R_i$ that one is derived from the other. It can be concluded that while \textsc{dsm-hd} is vulnerable to the reverse engineering attack when a coefficient is protected by multiple decoys, an adversary cannot gain any advantage over brute-forcing the key for \textsc{dsm-rd} and \textsc{dsm-hdrd}.

% It can be concluded that the security level of the DSMs against the reverse engineering attack is 0 for \cite{aksoy21} and \textsc{dsm-hd}, and $k$ for \textsc{dsm-rd} and \textsc{dsm-hdrd}.

% arch0 is hd
% arch0_tech3 is hd
% arch0_tech0 is random

%% file: results.tex
\section{Experimental Results}
\label{sec:results}

We used three FIR filters whose specifications are given in Table~\ref{tab:fir-specs}. These filters are obfuscated using the proposed technique. Moreover, their TMCM blocks, including the original coefficients, are locked by conventional methods~\cite{roy08, yasin16} and obfuscated by the technique of~\cite{aksoy21} using a Hamming distance based DSM similar to {\sc dsm-hd}. This section presents the synthesis results of encrypted filters, resiliency of encrypted TMCM blocks to reverse engineering and \mbox{oracle-less~\cite{alaql21}} attacks, and behavior of encrypted filters.

\begin{table}[t]
	\centering
	\scriptsize
	\caption{fir filter specifications.} 
	\vspace*{-3mm}
	\begin{tabular}{|c|c||c|c|c|c|c|c|}
		\hline
		Index & Type & $N$ & $w_p$ & $w_s$ & $\delta_p$ & $\delta_s$ & Q \\
		\hline \hline
		1 & low-pass  & 29  & 0.3   & 0.5   & 0.00316 & 0.00316 & 14 \\
		2 & low-pass  & 59  & 0.125 & 0.225 & 0.01    & 0.001   & 14 \\
		3 & high-pass & 105 & 0.8   & 0.7   & 0.005   & 0.001   & 14 \\
		\hline
	\end{tabular}
	\label{tab:fir-specs}
	\vspace*{-6mm}
\end{table}

\subsection{Synthesis of Encrypted Filters}

In order to find the hardware complexity of encrypted FIR filters, they are synthesized when the bit-width of the filter input is 32. Note that logic synthesis is performed by Cadence Genus using a commercial 65\,nm cell library. The encrypted designs are validated in simulation using 10,000 randomly generated inputs where the collected switching activity data is used to compute power dissipation. Table~\ref{tab:synth-fir} presents the gate-level synthesis results of filters encrypted using $p$ key inputs. In this table, \textit{area}, \textit{delay}, and \textit{power} stand for the total area in $\mu m^2$, delay in the critical path in $ps$, and total power dissipation in $\mu W$, respectively. Note that encrypted filters with minimum area are shown in bold. 

\begin{table}[t]
	\centering
	\scriptsize
	\vspace*{-1mm}
	\caption{Synthesis results of encrypted fir filters.}
	\vspace*{-2mm}
	\begin{tabular}{|@{\hskip3pt}c@{\hskip3pt}|@{\hskip3pt}c@{\hskip3pt}|@{\hskip3pt}l@{\hskip3pt}|l||c|c|c|}
		\hline
		\multirow{2}{*}{Index} & \multirow{2}{*}{$p$} & \multirow{2}{*}{Encryption} & \multirow{2}{*}{Technique} & \multicolumn{3}{c|}{Synthesis Results}\\
		\cline{5-7}
		& & & & area & delay & power\\
		\hline \hline
		\multirow{6}{*}{1} & \multirow{6}{*}{32}  & \multirow{2}{*}{Logic Locking} & \cite{roy08}                 & 14596 & 5708 & 1668 \\
		&                      &                                & \textbf{\cite{yasin16}}      & \textbf{14452} & \textbf{5423} & \textbf{1428} \\
		\cline{3-7}                                             
		&                      & \multirow{4}{*}{Obfuscation}   & \cite{aksoy21}               & 14483 & 6599 & 1545 \\
		&                      &                                & {\sc ours - dsm-hd}          & 14475 & 6123 & 1523 \\
		&                      &                                & {\sc ours - dsm-rd}          & 14818 & 6625 & 1672 \\
		&                      &                                & {\sc ours - dsm-hdrd}        & 14616 & 5753 & 1555 \\
		\hline \hline                                                              
		\multirow{6}{*}{2} & \multirow{6}{*}{64}  & \multirow{2}{*}{Logic Locking} & \cite{roy08}                 & 26096 & 6075 & 2383 \\
		&                      &                                & \cite{yasin16}               & 25898 & 6400 & 2229 \\
		\cline{3-7}                                             
		&                      & \multirow{4}{*}{Obfuscation}   & \cite{aksoy21}               & 25856 & 6764 & 2266 \\
		&                      &                                & {\sc \textbf{ours - dsm-hd}} & \textbf{25851} & \textbf{6523} & \textbf{2202} \\
		&                      &                                & {\sc ours - dsm-rd}          & 26146 & 6598 & 2248 \\
		&                      &                                & {\sc ours - dsm-hdrd}        & 25949 & 6563 & 2249 \\
		\hline \hline                                                              
		\multirow{6}{*}{3} & \multirow{6}{*}{128} & \multirow{2}{*}{Logic Locking} & \cite{roy08}                 & 45466 & 7072 & 3979 \\
		&                      &                                & \cite{yasin16}               & 45404 & 7179 & 3840 \\
		\cline{3-7}                                             
		&                      & \multirow{4}{*}{Obfuscation}   & \textbf{\cite{aksoy21}}      & \textbf{45387} & \textbf{7321} & \textbf{3882} \\
		&                      &                                & {\sc ours - dsm-hd}          & 45437 & 6976 & 3837 \\
		&                      &                                & {\sc ours - dsm-rd}          & 46059 & 6458 & 3901 \\
		&                      &                                & {\sc ours - dsm-hdrd}        & 45690 & 7383 & 3927 \\
		\hline
	\end{tabular}
	\label{tab:synth-fir}
	\vspace*{-4mm}
\end{table}

Observe from Table~\ref{tab:synth-fir} that among the logic locking methods, the method of~\cite{yasin16} leads to locked filters with the smallest area, achieving a maximum of 2.71\% gain in area with respect to the method of~\cite{roy08}. Also, among the obfuscation techniques, the technique of~\cite{aksoy21} and the proposed technique using \mbox{\sc dsm-hd} leads to obfuscated filters with the smallest area, while the filters obfuscated by the proposed technique using \mbox{\sc dsm-rd} have the largest area. In this case, the gain in area reaches up to 2.31\% when compared to the proposed technique using \mbox{\sc dsm-rd}. This is because while \mbox{\sc dsm-hd} uses decoys whose Hamming distance to the filter coefficients is the smallest, \mbox{\sc dsm-rd} favors random decoys. Also, the proposed technique using \mbox{\sc dsm-hd} and \mbox{\sc dsm-hdrd} generates filters with hardware complexity very close to each other, since \mbox{\sc dsm-hdrd} uses decoys, whose Hamming distance to each coefficient is the smallest, for coefficients protected by a single decoy. Note also that there exist obfuscated filters that occupy less area than locked filters, e.g., Filters 2 and 3.

\begin{figure}[t]
	\centering
	%\vspace*{-1mm}
	\includegraphics[width=9cm]{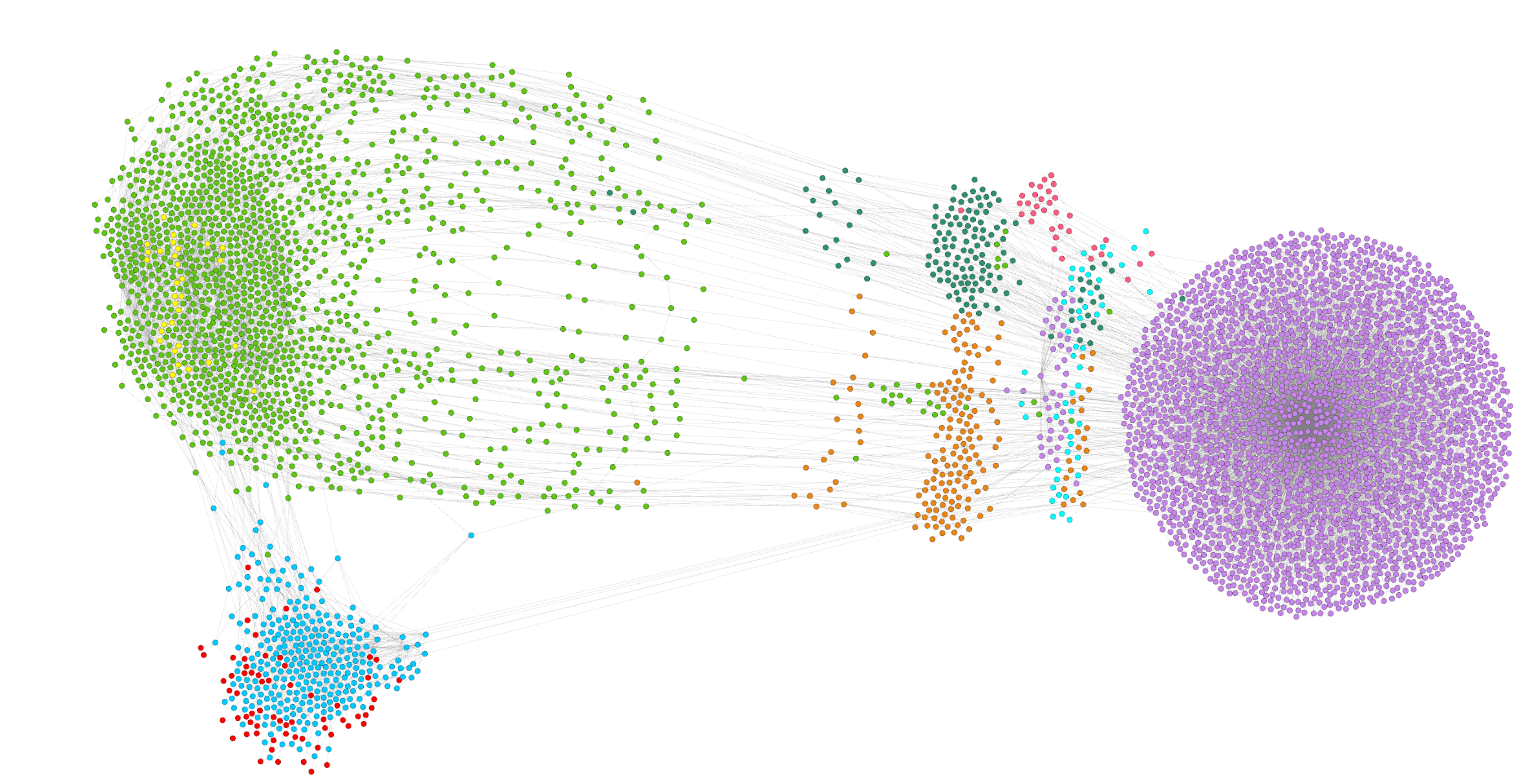}
	\vspace*{-8mm}
	\caption{Netlist graph of identified blocks in the obfuscated filter.}
	\vspace*{-6mm}
	\label{fig:reveng-fir02}
\end{figure}

\subsection{Attacks to Encrypted TMCM Blocks}

\subsubsection{Reverse Engineering}

Fig.~\ref{fig:reveng-fir02} shows the blocks of Filter 2 obfuscated by the proposed technique using \mbox{\sc dsm-hd} which are identified by the reverse engineering method of~\cite{baeh2020}. Note that while the yellow and red dots denote the filter inputs and key inputs, respectively, the cyan dots denote the filter outputs. Also, the bright and light green dots represent the multiplier and adder, respectively. The other dots denote the glue logic.

Table~\ref{tab:reveng-tmcm} shows the results on the reverse engineering attack, in which $acc$ is the probability that the design is obfuscated by a Hamming distance based DSM, as reported by the XGBClassifier, $vc$ is the number of vulnerable coefficients, actually the number of coefficients protected by multiple decoys, and $cdc$ is the number of correctly determined coefficients found by the \mbox{\sc doc-hd} script described in Section~\ref{subsec:foc}. Note that this script is not applicable (NA) to DSMs with an $acc$ value close to 0. Finally, $apc$ is the number of all possible combinations of constants to be tried to find coefficients.

Observe from Table~\ref{tab:reveng-tmcm} that a Hamming distance based DSM can be found with a high accuracy, except \mbox{\sc dsm-rd} on Filter 2. The vulnerable coefficients are easily identified in TMCM blocks obfuscated by the technique of~\cite{aksoy21} and the proposed technique using {\sc dsm-hd}, reducing the effort of the attacker on exhaustive search of filter coefficients. As {\sc dsm-hdrd} selects a single decoy for a coefficient using {\sc dsm-hd}, it is correctly classified as a Hamming distance based DSM. Still, \mbox{\sc doc-hd} fails to identify the coefficients, as {\sc dsm-hdrd} selects these multiple decoys randomly for a coefficient using {\sc dsm-rd}.

\begin{table}[t]
	\centering
	\scriptsize
	\vspace*{-1mm}
	\caption{Results of the reverse engineering attack on obfuscated tmcm blocks.}
	\vspace*{-2mm}
	\begin{tabular}{|c|c|l||c|c|c|c|}
		\hline
		\multirow{2}{*}{Index} & \multirow{2}{*}{$p$} & \multirow{2}{*}{Technique} & \multicolumn{4}{c|}{Attack Results}\\
		\cline{4-7}
		& & & acc & vc & cdc & apc\\
		\hline \hline
		\multirow{4}{*}{1} & \multirow{4}{*}{32}   & \cite{aksoy21}        & 0.9986 & 3  & 3  & $2^{26}$ \\
		&                       & {\sc ours - dsm-hd}   & 0.8896 & 3  & 3  & $2^{26}$ \\
		&                       & {\sc ours - dsm-rd}   & 0.0086 & NA & NA & $2^{32}$ \\
		&                       & {\sc ours - dsm-hdrd} & 0.9869 & 3  & 0  & $2^{32}$ \\
		\hline \hline                                                                     
		\multirow{4}{*}{2} & \multirow{4}{*}{64}   & \cite{aksoy21}        & 0.9999 & 5  & 5  & $2^{54}$ \\
		&                       & {\sc ours - dsm-hd}   & 0.9999 & 5  & 5  & $2^{54}$ \\
		&                       & {\sc ours - dsm-rd}   & 0.9991 & 5  & 0  & $2^{64}$ \\
		&                       & {\sc ours - dsm-hdrd} & 0.9998 & 5  & 0  & $2^{64}$ \\
		\hline \hline                                                                     
		\multirow{4}{*}{3} & \multirow{4}{*}{128}  & \cite{aksoy21}        & 0.9999 & 23 & 23 & $2^{82}$ \\
		&                       & {\sc ours - dsm-hd}   & 0.9999 & 23 & 23 & $2^{82}$ \\
		&                       & {\sc ours - dsm-rd}   & 0.0009 & NA & NA & $2^{128}$ \\
		&                       & {\sc ours - dsm-hdrd} & 0.9983 & 23 & 0  & $2^{128}$ \\
		\hline
	\end{tabular}
	\label{tab:reveng-tmcm}
	\vspace*{-4mm}
\end{table}

\begin{table}[t]
	\centering
	\scriptsize
	\caption{Results of the oracle-less attack on encrypted tmcm blocks.}
	\vspace*{-2mm}
	\begin{tabular}{|c|c|l|l||c|c|c|}
		\hline
		\multirow{2}{*}{Index} & \multirow{2}{*}{$p$} & \multirow{2}{*}{Encryption} & \multirow{2}{*}{Technique} & \multicolumn{3}{c|}{Attack Results}\\
		\cline{5-7}
		& & & & dk & pcdk & time\\
		\hline \hline
		\multirow{6}{*}{1} & \multirow{6}{*}{32}  & \multirow{2}{*}{Logic Locking} & \cite{roy08}          & 3  & 100 & 5.35 \\
		&                      &                                & \cite{yasin16}        & 32 & 100 & 4.94 \\
		\cline{3-7}                                                                
		&                      & \multirow{4}{*}{Obfuscation}   & \cite{aksoy21}        & 32 & 53.1 & 5.27 \\
		&                      &                                & {\sc ours - dsm-hd}   & 32 & 62.5 & 5.31 \\
		&                      &                                & {\sc ours - dsm-rd}   & 32 & 62.6 & 3.32 \\
		&                      &                                & {\sc ours - dsm-hdrd} & 32 & 50   & 6.04 \\
		\hline \hline                                                              
		\multirow{6}{*}{2} & \multirow{6}{*}{64}  & \multirow{2}{*}{Logic Locking} & \cite{roy08}          & 8  & 50  & 9.27  \\
		&                      &                                & \cite{yasin16}        & 64 & 100 & 10.07 \\
		\cline{3-7}                                                                
		&                      & \multirow{4}{*}{Obfuscation}   & \cite{aksoy21}        & 64 & 51.6 & 9.45 \\
		&                      &                                & {\sc ours - dsm-hd}   & 64 & 59.3 & 9.73 \\
		&                      &                                & {\sc ours - dsm-rd}   & 63 & 57.1 & 6.59 \\
		&                      &                                & {\sc ours - dsm-hdrd} & 64 & 53.1 & 10.08 \\
		\hline \hline                                                              
		\multirow{6}{*}{3} & \multirow{6}{*}{128} & \multirow{2}{*}{Logic Locking} & \cite{roy08}          & 19  & 68.4 & 21.61 \\
		&                      &                                & \cite{yasin16}        & 128 & 100  & 23.43 \\
		\cline{3-7}                                                                
		&                      & \multirow{4}{*}{Obfuscation}   & \cite{aksoy21}        & 128 & 63.2 & 22.11 \\
		&                      &                                & {\sc ours - dsm-hd}   & 128 & 63.2 & 21.1  \\
		&                      &                                & {\sc ours - dsm-rd}   & 128 & 64   & 16    \\
		&                      &                                & {\sc ours - dsm-hdrd} & 128 & 50   & 27.76 \\
		\hline
	\end{tabular}
	\label{tab:scope-tmcm}
	\vspace*{-6mm}
\end{table}

\subsubsection{Oracle-less Attack}

The attack of~\cite{alaql21} is used to find the secret key of encrypted TMCM blocks. Table~\ref{tab:scope-tmcm} presents its solutions, where \textit{dk} and \textit{pcdk} stand for the number of deciphered key inputs and the percentage of correct deciphered key bits, respectively and \textit{time} denotes its run-time in seconds.

Observe from Table~\ref{tab:scope-tmcm} that the oracle-less attack can decipher a small number of key inputs of designs locked by the technique of~\cite{roy08} with a high accuracy, except Filter 2. It can determine the secret key of designs locked by the technique of~\cite{yasin16}. In contrast, its guesses on the key inputs of obfuscated designs are slightly better than a random guess.

\subsection{Filter Behavior}

\begin{figure*}[t]
	\centering
	\vspace*{-2mm}
	\parbox{5.8cm}{\centerline{\includegraphics[width=6.5cm]{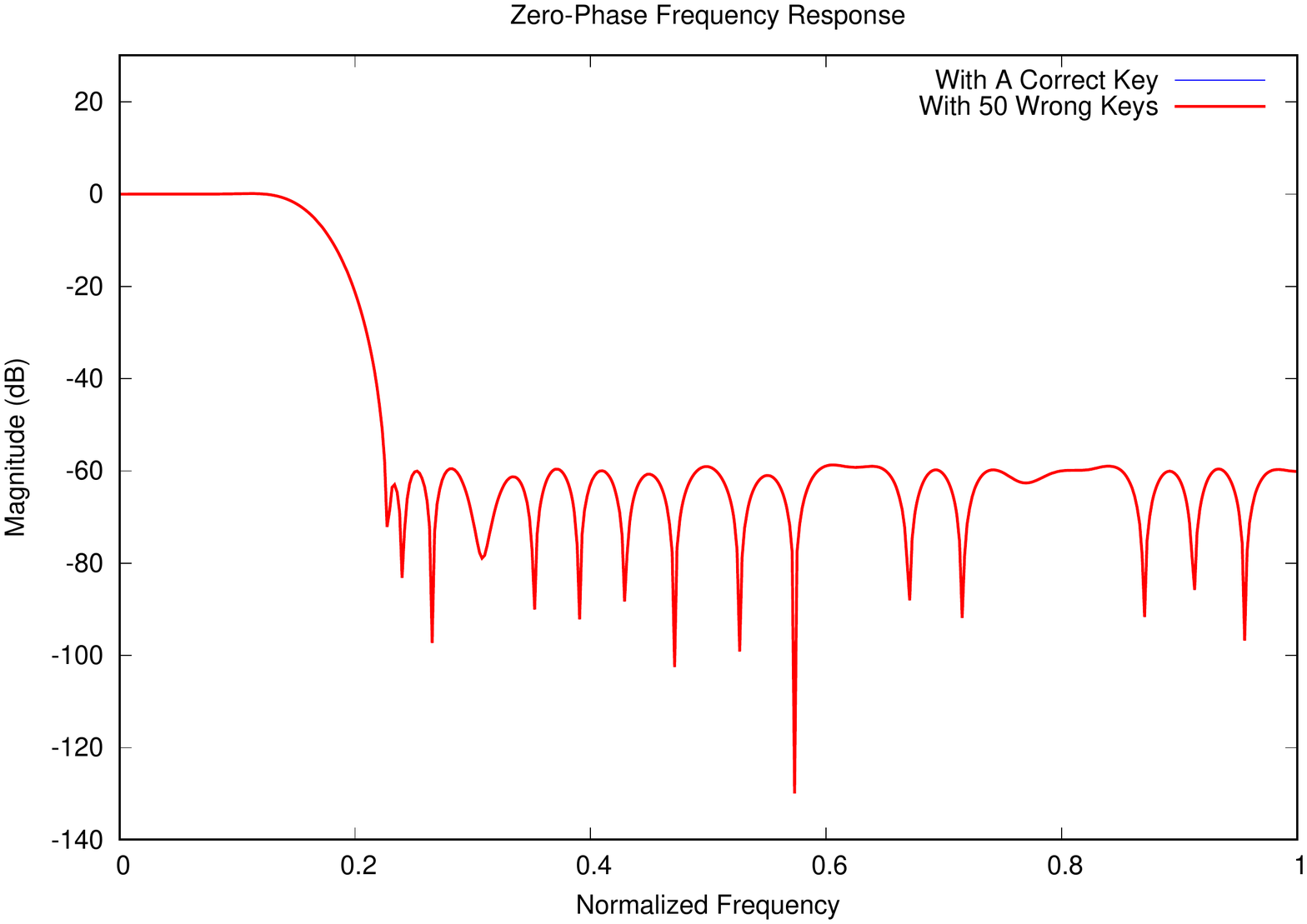}}}\
	\parbox{5.8cm}{\centerline{\includegraphics[width=6.5cm]{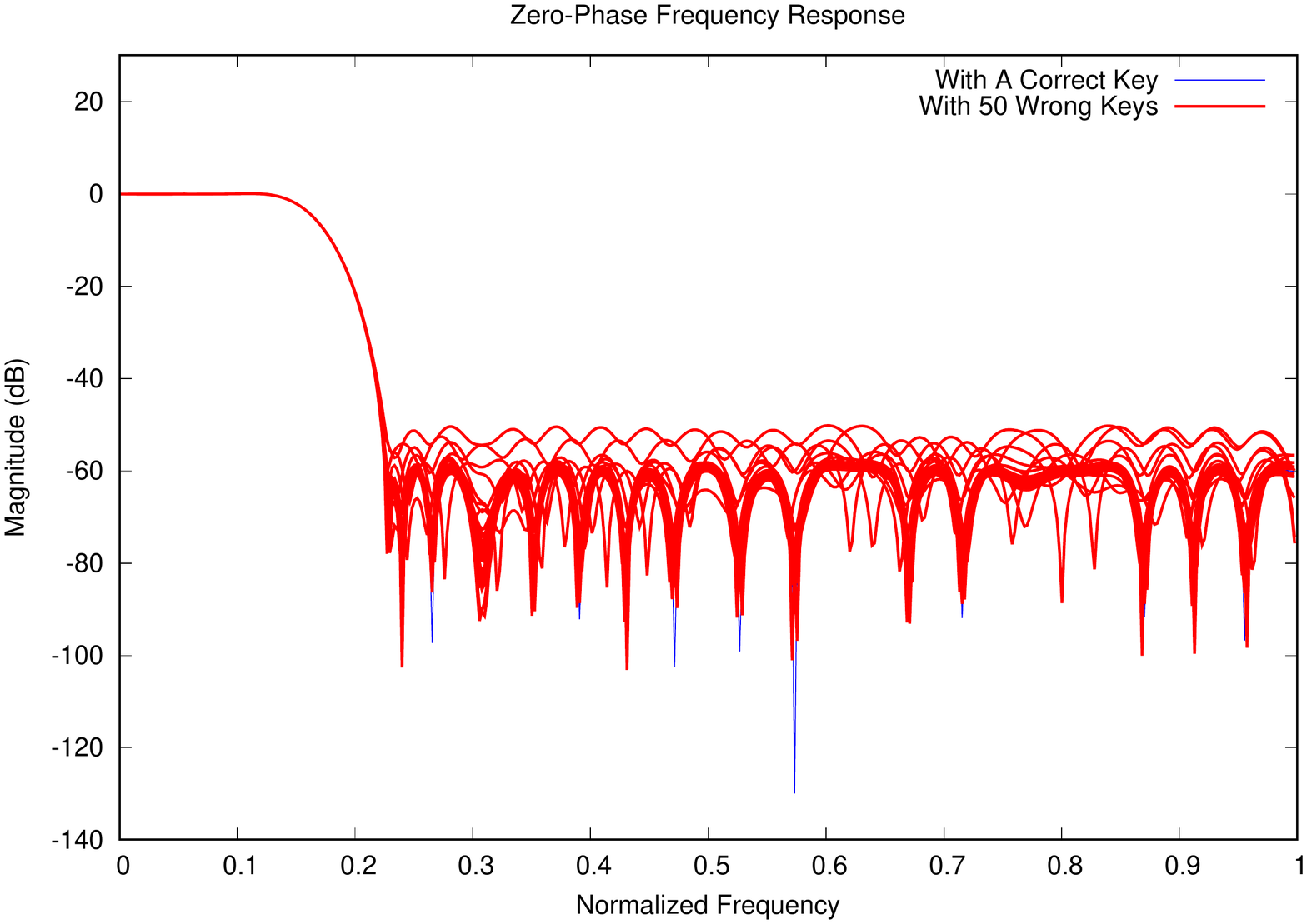}}}\
	\parbox{5.8cm}{\centerline{\includegraphics[width=6.5cm]{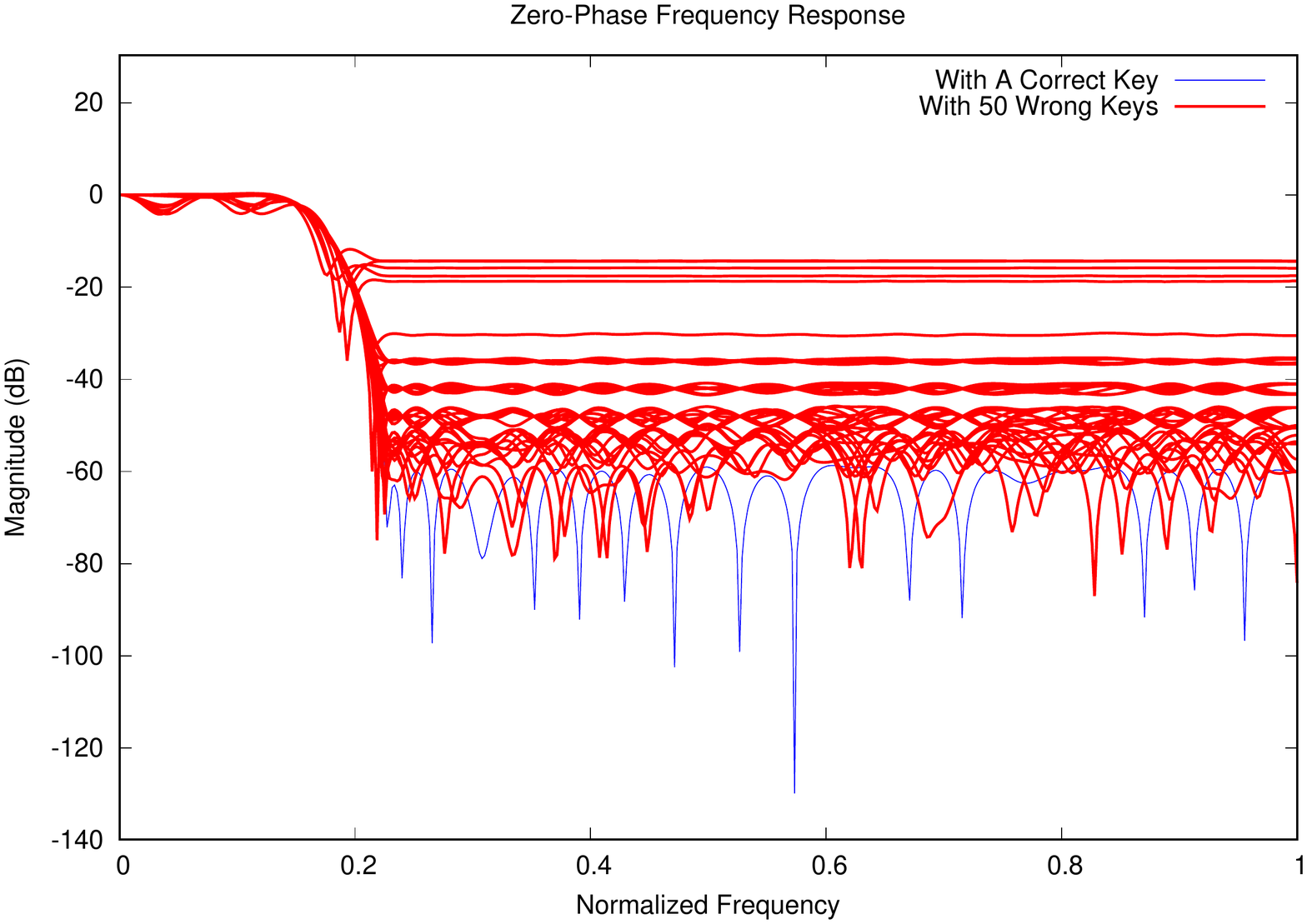}}}\
	
	\vspace*{-3mm}
	
	\parbox{5.8cm}{\centerline{\scriptsize (a)}}\
	\parbox{5.8cm}{\centerline{\scriptsize (b)}}\
	\parbox{5.8cm}{\centerline{\scriptsize (c)}}\
	\vspace*{-3mm}
	\caption{ZPFRs of FIR filters: (a)~locked by the method of~\cite{yasin16}; (b)~obfuscated by the method of~\cite{aksoy21}; (c)~obfuscated by the proposed method using {\sc dsm-hdrd}.}  
	\label{fig:zpfr-enc}
	\vspace*{-6mm}
\end{figure*}

In order to find the behavior of an encrypted FIR filter, Filter 2 is considered. The ZPFR of an encrypted filter given in Eq.~\ref{eqn:zpfr} is computed based on the filter coefficients determined as given in Eq.~\ref{eqn:coefs} after applying a specific filter input and observing the filter output. Fig.~\ref{fig:zpfr-enc} presents ZPFRs of filters locked by the method of~\cite{yasin16}, obfuscated by the technique of~\cite{aksoy21} and the proposed technique using {\sc dsm-hdrd}. The filter behavior is obtained using 50 wrong keys, which are very close to the secret key in terms of the Hamming distance.

%Moreover, Fig.~\ref{fig:zpfr-firobf} shows ZPFRs of filters obfuscated by the proposed technique. 

Observe from Fig.~\ref{fig:zpfr-enc} that while the filter locked by the method of~\cite{yasin16} has the original filter behavior under the given wrong keys, the filter obfuscated by the technique of~\cite{aksoy21} has a behavior very similar to the original one under some wrong keys. In contrast, the filter obfuscated by our proposed technique exhibits incorrect behavior under every wrong key. We note that this property is also observed on filters obfuscated by the proposed technique using {\sc dsm-hd} and {\sc dsm-rd}. Observe from Tables~\ref{tab:synth-fir}-\ref{tab:scope-tmcm} that the proposed technique using {\sc dsm-hdrd} generates filters with competitive hardware complexity and high security.

%%% OlD MAteRIaL %%%
\comment{
	\begin{figure*}[t]
		\centering
		\parbox{5.8cm}{\centerline{\includegraphics[width=6.5cm]{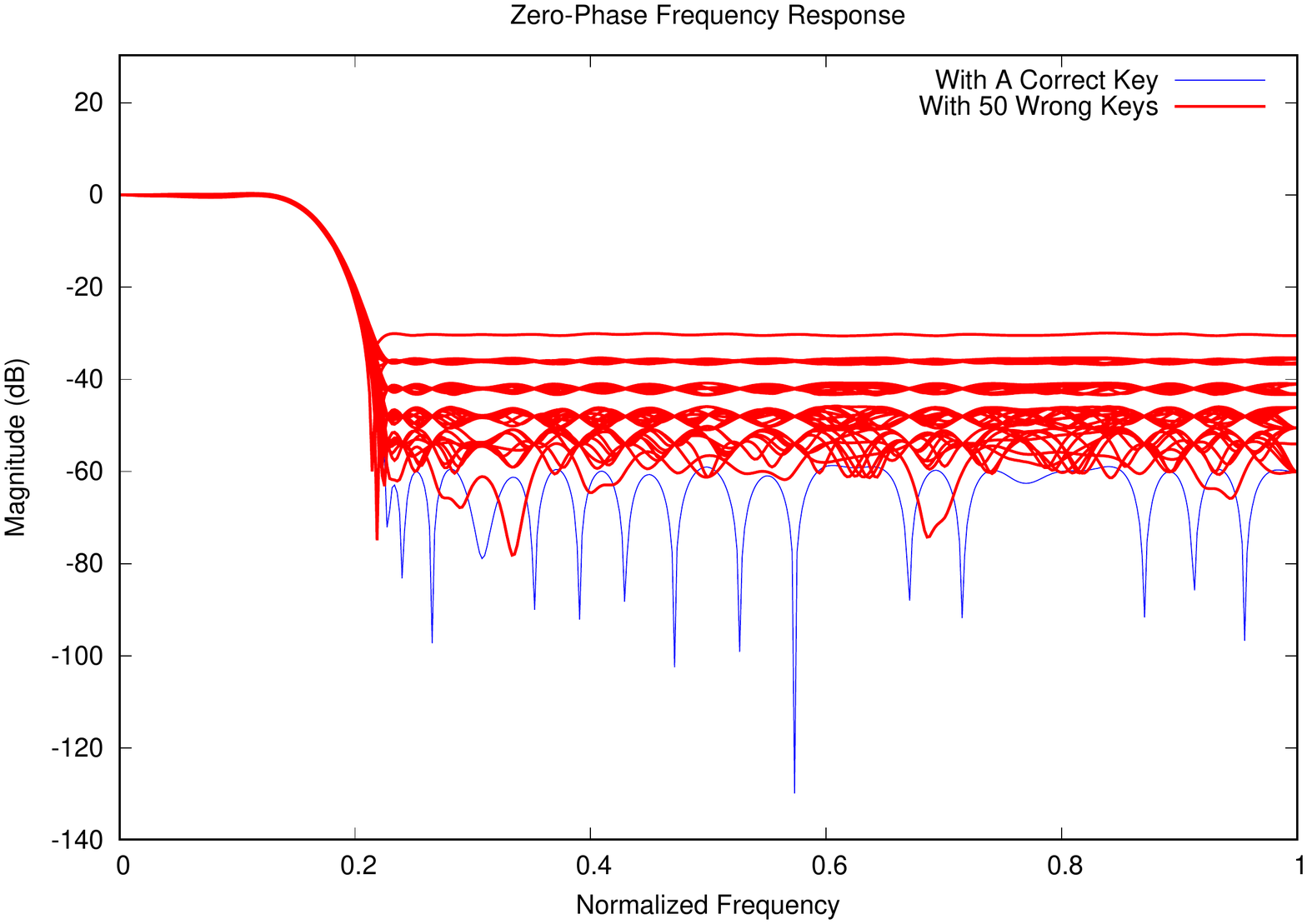}}}\
		\parbox{5.8cm}{\centerline{\includegraphics[width=6.5cm]{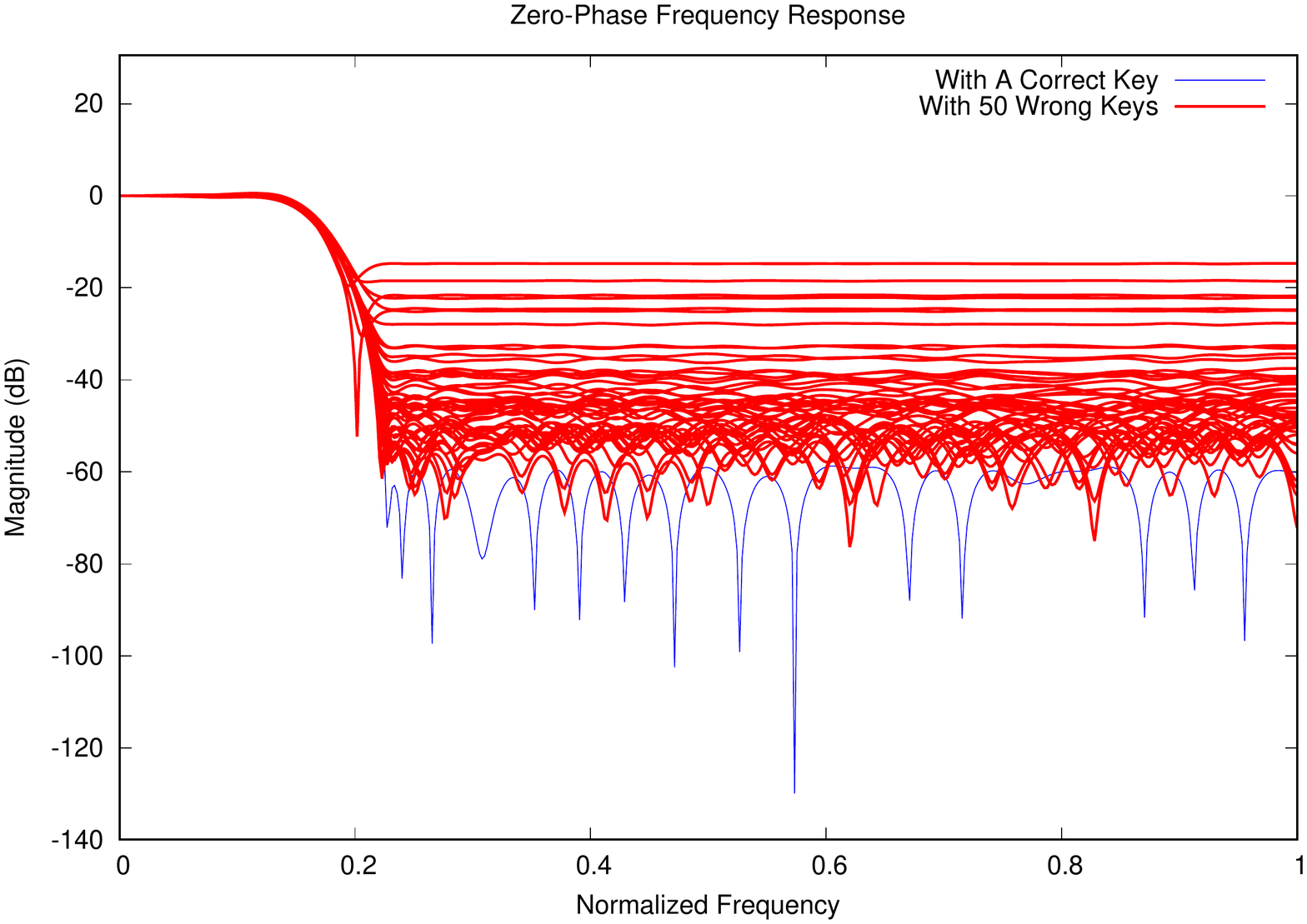}}}\
		\parbox{5.8cm}{\centerline{\includegraphics[width=6.5cm]{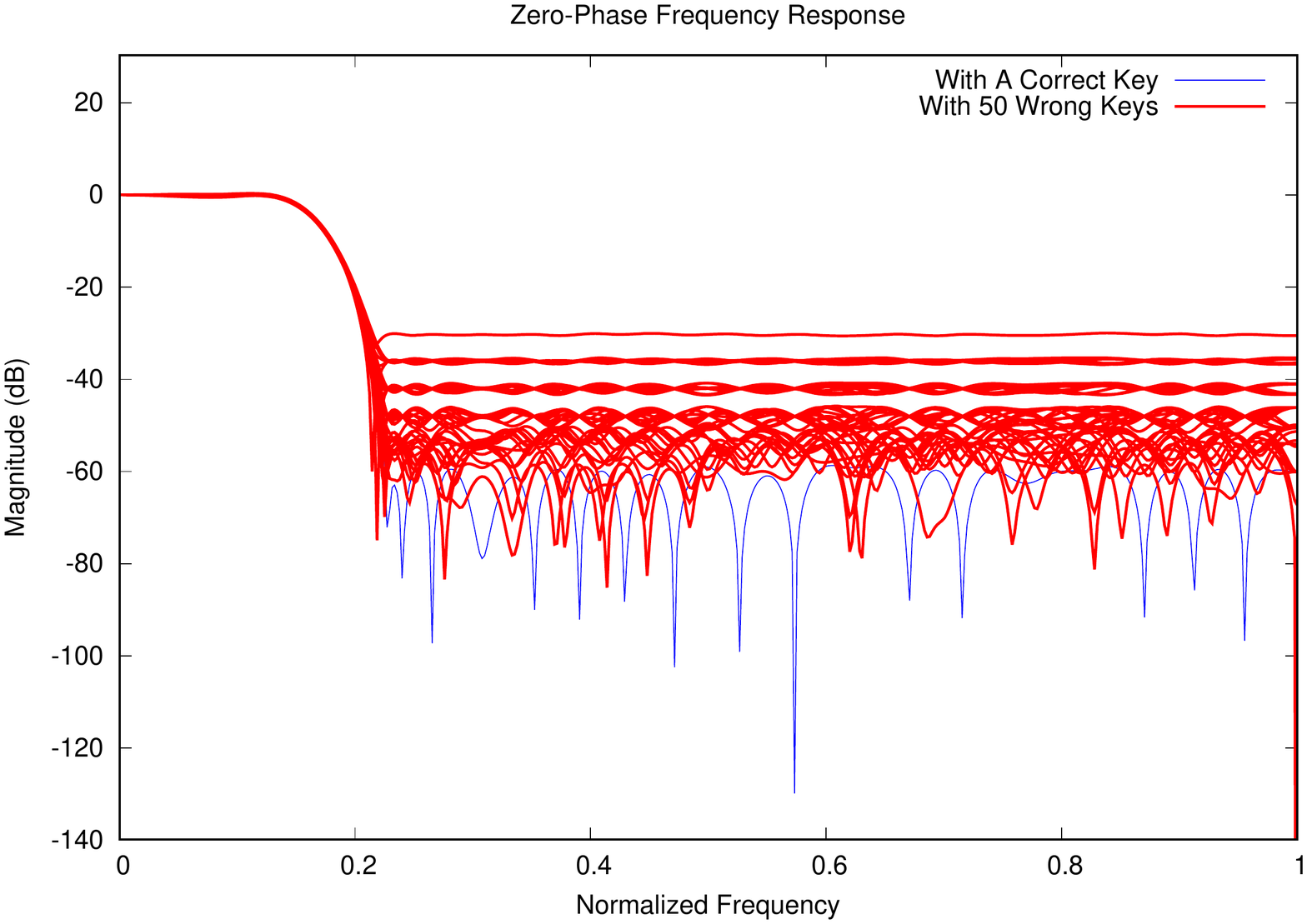}}}\
		
		\vspace*{-3mm}
		
		\parbox{5.8cm}{\centerline{\scriptsize (a)}}\
		\parbox{5.8cm}{\centerline{\scriptsize (b)}}\
		\parbox{5.8cm}{\centerline{\scriptsize (c)}}\
		\vspace*{-3mm}
		\caption{ZPFRs of FIR filters obfuscated by the proposed technique: (a)~using {\sc dsm1}; (b)~using {\sc dsm2}; (c)~using {\sc dsm3}.}  
		\label{fig:zpfr-firobf}
		\vspace*{-6mm}
	\end{figure*}

	\begin{figure}[t]
		\centering
		\vspace*{-4mm}
		\includegraphics[width=9cm]{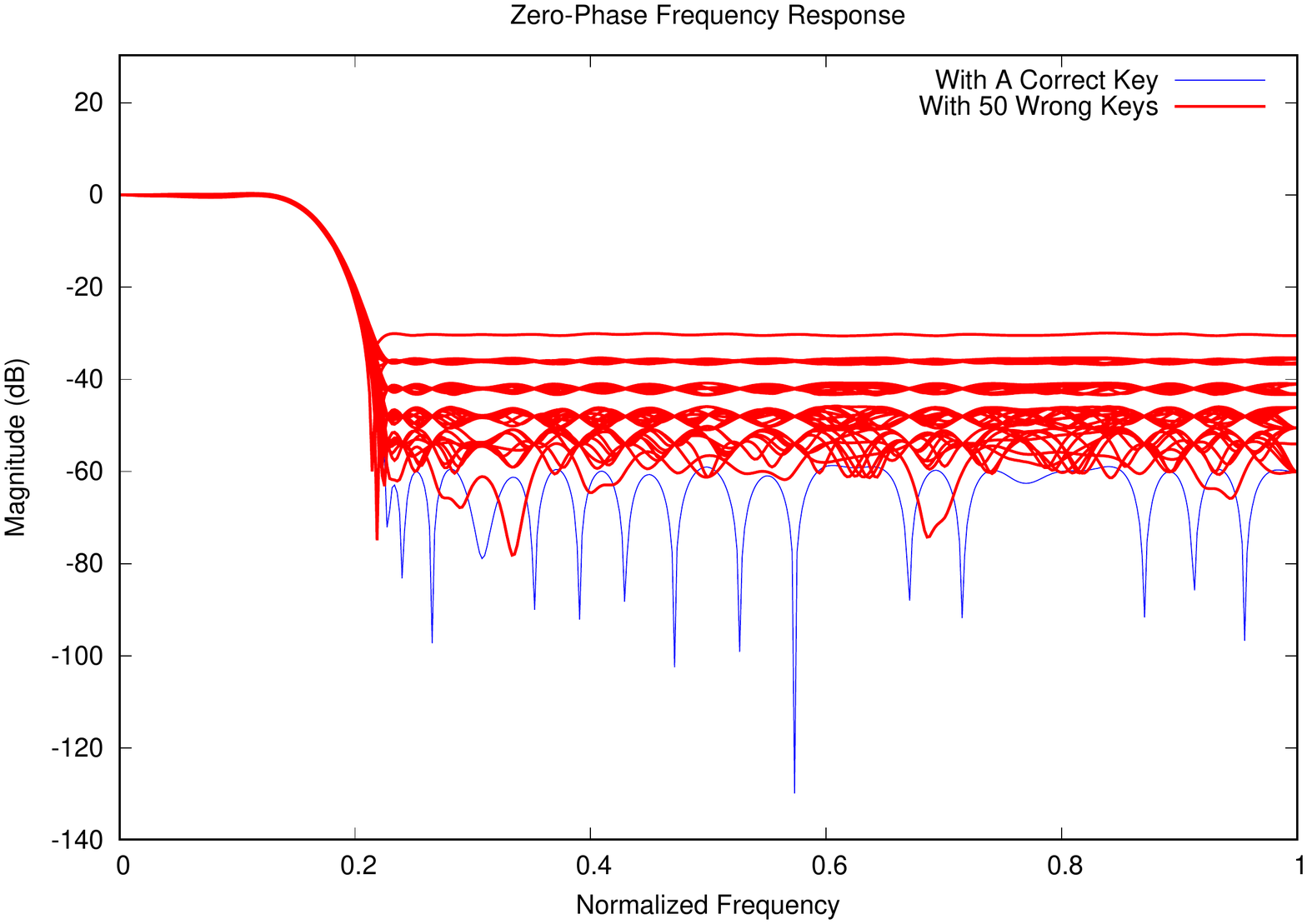}
		\vspace*{-12mm}
		\caption{The ZPFR of the FIR filter obfuscated by the proposed technique.}
		\vspace*{-6mm}
		\label{fig:zpfr-firobf}
	\end{figure}
	
	\subsection{Synthesis of Encrypted TMCM Blocks}
	\label{subsec:enc-tmcm}
	
	In order to find the hardware complexity of the encrypted TMCM blocks, they are synthesized when the bit-width of the filter input is 32. Note that logic synthesis is performed by Cadence Genus using a commercial 65\,nm cell library. The encrypted designs are validated in simulation using 10,000 randomly generated inputs where the collected switching activity data is used to compute power dissipation. Table~\ref{tab:synth-tmcm} presents the gate-level synthesis results of encrypted TMCM blocks of filters. In this table, \textit{area}, \textit{delay}, and \textit{power} stand for the total area in $\mu m^2$, delay in the critical path in $ps$, and total power dissipation in $\mu W$, respectively. Note that TMCM designs with minimum area are shown in bold.
	
	\begin{table}[t]
		\centering
		\scriptsize
		\caption{Synthesis results of encrypted tmcm blocks.}
		\vspace*{-2mm}
		\begin{tabular}{|c|c|c|l||c|c|c|}
			\hline
			\multirow{2}{*}{Index} & \multirow{2}{*}{$p$} & \multirow{2}{*}{Encryption} & \multirow{2}{*}{Technique} & \multicolumn{3}{c|}{Synthesis Results}\\
			\cline{5-7}
			& & & & area & delay & power\\
			\hline \hline
			\multirow{6}{*}{1} & \multirow{6}{*}{32}  & \multirow{2}{*}{Logic Locking} &  \cite{roy08}            & 3429 & 5287 & 1435 \\
			&                      &                                &  \textbf{\cite{yasin16}} & \textbf{3336} & \textbf{4808} & \textbf{1250} \\
			\cline{3-7}
			&                      & \multirow{4}{*}{Obfuscation}   & \cite{aksoy21}           & 3427 & 5776 & 1416 \\
			&                      &                                & {\sc ours - dsm1}        & 3528 & 4123 & 1383 \\
			&                      &                                & {\sc ours - dsm2}        & 3610 & 5647 & 1558 \\
			&                      &                                & {\sc ours - dsm3}        & 3434 & 5792 & 1466 \\
			\hline \hline
			\multirow{6}{*}{2} & \multirow{6}{*}{64}  & \multirow{2}{*}{Logic Locking} & \cite{roy08}             & 3656 & 5887 & 1685 \\
			&                      &                                & \cite{yasin16}           & 3650 & 5702 & 1518 \\
			\cline{3-7}                                                                  
			&                      & \multirow{4}{*}{Obfuscation}   & \textbf{\cite{aksoy21}}  & \textbf{3592} & \textbf{5970} & \textbf{1536} \\
			&                      &                                & {\sc ours - dsm1}        & 3609 & 5739 & 1555 \\
			&                      &                                & {\sc ours - dsm2}        & 3959 & 6096 & 1724 \\
			&                      &                                & {\sc ours - dsm3}        & 3653 & 4058 & 1582 \\
			\hline \hline
			\multirow{6}{*}{3} & \multirow{6}{*}{128} & \multirow{2}{*}{Logic Locking} & \cite{roy08}             & 4003 & 6452 & 2000 \\
			&                      &                                & \textbf{\cite{yasin16}}  & \textbf{3936} & \textbf{6088} & \textbf{1733} \\
			\cline{3-7}                                                                  
			&                      & \multirow{4}{*}{Obfuscation}   & \cite{aksoy21}           & 4124 & 5909 & 1706 \\
			&                      &                                & {\sc ours - dsm1}        & 4149 & 5928 & 1828 \\
			&                      &                                & {\sc ours - dsm2}        & 4537 & 7031 & 2416 \\
			&                      &                                & {\sc ours - dsm3}        & 4183 & 6063 & 1805 \\
			\hline \hline
			\multirow{6}{*}{4} & \multirow{6}{*}{128} & \multirow{2}{*}{Logic Locking} & \cite{roy08}             & 4303 & 6142 & 2075 \\
			&                      &                                & \cite{yasin16}           & 4350 & 6265 & 1996 \\
			\cline{3-7}                                                                    
			&                      & \multirow{4}{*}{Obfuscation}   & \textbf{\cite{aksoy21}}  & \textbf{4299} & \textbf{6030} & \textbf{2066} \\
			&                      &                                & {\sc ours - dsm1}        & 4326 & 6151 & 2101 \\
			&                      &                                & {\sc ours - dsm2}        & 5028 & 6425 & 2332 \\
			&                      &                                & {\sc ours - dsm3}        & 4482 & 6320 & 2054 \\
			\hline
		\end{tabular}
		\label{tab:synth-tmcm}
		\vspace*{-6mm}
	\end{table}
	
	Observe from Table~\ref{tab:synth-tmcm} that among the logic locking methods, the method of~\cite{yasin16} generally leads to a design with the smallest area, where its gain reaches up to 2.71\% with respect to the method of~\cite{roy08}. Among the obfuscation techniques, the technique of~\cite{aksoy21} generally leads to a design with the smallest area, where its gain reaches up to 2.86\% with respect to the proposed technique. This is simply because the technique of~\cite{aksoy21} uses decoys that are very close to the filter coefficients in terms of the Hamming distance which enables the sharing of more common operations. Observe also that an obfuscation technique can lead to a design with less area when compared to locked designs, e.g., Filters 2 and 4.
}

%% file: conclusions.tex
\section{Conclusions}
\label{sec:conclusions}

This paper presented a filter design technique, which can obfuscate an FIR filter such that it exhibits a correct behavior only when the secret key is provided. It also introduced a reverse engineering technique to determine the original filter coefficients through selected decoys. The proposed obfuscation technique was compared to conventional logic locking methods and a recently proposed obfuscation technique. It was shown that it can generate filters whose hardware complexity is very close to those of filters encrypted by previously proposed techniques. However, filters obfuscated by our proposed technique present better security properties.  

%Otherwise, its behavior is out of the filter specification. The proposed technique uses decoy(s) to obfuscate each original filter coefficient, which are taken beyond the lower and upper bounds of the filter coefficient. 

%% file: acknowledgment.tex
\section*{Acknowledgment}

This work was partially supported by the EC through the European Social Fund in the context of the project ``ICT programme''. It was also partially supported by European Union's Horizon 2020 research and innovation programme under grant agreement No 952252 (SAFEST).